\documentstyle[12pt]{article}
  \advance\oddsidemargin  by -1in
  \advance\evensidemargin by -1in
\def\lsim{\mathrel{\rlap{\lower4pt\hbox{\hskip1pt$\sim$}}
    \raise1pt\hbox{$<$}}}         
\def\gsim{\mathrel{\rlap{\lower4pt\hbox{\hskip1pt$\sim$}}
    \raise1pt\hbox{$>$}}}         

\def\be{\begin{equation}}
\def\ee{\end{equation}}
\def\bq{\begin{eqnarray}}
\def\eq{\end{eqnarray}}

\begin{document}
\pagestyle{empty}
\hfill{DFTT 88/95}

\hfill{LYCEN 9544}

\hfill{December 1995}

\vspace{2.0cm}
 
\begin{center}
 
{\large \bf HEAVY QUARKS IN CHARGED--CURRENT \\ 
DEEP INELASTIC SCATTERING : \\ A SYSTEMATIC PHENOMENOLOGICAL STUDY \\
}

\vspace{1.0cm}

{\large Vincenzo Barone$^{a,b}$ 
and Marco Genovese$^{a,c}$\\}

\vspace{1.0cm} 

{\it $^{a}$Dipartimento di
Fisica Teorica dell'Universit\`a \\
and INFN, Sezione di
Torino,   I--10125 Torino, Italy 
\medskip\\
$^{b}$II Facolt{\`a} di Scienze MFN, I--15100 Alessandria, Italy
\medskip\\
$^{c}$Institut de Physique Nucl\'eaire de Lyon \\ Universit\'e
Claude Bernard,
F-69622 Villeurbanne Cedex, France}

\vspace{1.0cm}

{\large \bf Abstract \bigskip\\ }

\end{center}

We present a systematic QCD analysis of the strange--charm and 
bottom--top contributions to transverse and 
longitudinal structure functions in charged--current deep inelastic
scattering. Various ${\cal O}(\alpha_s^1)$ schemes
are studied and compared. The dependence of their predictions
on the factorization scale $\mu^2$ is investigated.
The theoretical uncertainties resulting 
from the choice of the scheme and of $\mu^2$
are estimated.

\vskip 3cm

\vfill
 
\pagebreak

\baselineskip 16 pt
\pagestyle{plain}

Neutrino Deep Inelastic Scattering ($\nu$DIS) is an invaluable source
of information on nucleon parton densities. 
In particular, neutrino reactions play a fundamental
r\^ole in 
the determination of the strange
quark distribution \cite{BGNPZ5}.

Both to theorists and to experimentalists neutrino--induced deep 
inelastic processes  pose a number of 
subtle problems, due to the peculiar structure
of the weak currents. These currents are not conserved, 
hence large longitudinal contributions are expected to exist 
\cite{BGNPZ2},
which badly
break the Callan--Gross relation. Moreover, 
in charged--current weak DIS, 
flavours of different masses are mixed and 
interesting threshold effects occur \cite{BGNPZ1}. For instance, 
 the excitation of strangeness (a light flavour) produces charm (a heavy
flavour) via the lowest order transition $W^{+}s \rightarrow c$, and 
there is the simultaneous  
production of $\bar s$ and
$c$ in the higher--order $W$--gluon fusion 
reaction $W^{+}g \rightarrow \bar s c$. 
A similar situation takes place in the bottom--top
sector. 

Two important questions arise in charged--current DIS. 
They 
possess a great theoretical significance and, at the same time, 
 are constantly faced 
by experimentalists in their analyses. 

The first question is whether a leading order (LO) \footnote{We 
shall refer to 
the ${\cal O}(\alpha_s^0)$ 
quark excitation processes as ``leading order'', and to 
the ${\cal O}(\alpha_s^1)$ $W$-gluon fusion processes as 
``next-to-leading order''. This terminology is admittedly 
ambiguous and is used here only for practical purposes 
and to adhere to some experimental papers \cite{CCFR1,CCFR2}; 
it differs
from that adopted by other authors, {\it e.g.} \cite{GR}.}
approach, 
based on the quark excitation processes $W^{+}q' \rightarrow q, \,\,
W^{+}\bar q \rightarrow \bar q'$ ($q'=s,b;\,\, q=c,t$), and on 
the slow rescaling mechanism (see below), 
is sufficient to carry out a reliable 
data analysis. The slow rescaling (SR) procedure is appealing because 
it is relatively easy to perform, but it is known
to give unreliable results, in some instance. 
Previous theoretical \cite{BGNPZ5,BGNPZ2,BGNPZ1,BGNPZ3,BGNPZ4} 
and experimental \cite{CCFR1,CCFR2}
studies on the determination 
of the strange density from neutrino DIS 
have indeed taught us the 
importance of quark-mass corrections and current non-conservation
effects, which manifest themselves through the next--to--leading 
order (NLO)
$W$--gluon fusion diagrams. 
However, when heavy quarks are involved, there is no unique way
to take higher order effects into account in a partonic language. 
Various approaches exist, which hopefully (but not necessarily) 
are equivalent to each other.
 
Here comes the second problem announced above: the treatment 
of heavy flavours. 
It seems natural to make the notion of ``heavy'' and ``light'' 
depend on the physical scale of the process (this has
been emphasized in \cite{Tung}). At $Q^2 \simeq 10-30$ 
GeV$^2$, which is the typical average 
scale of present neutrino experiments, $s$ is light whereas $c$ is 
undoubtedly heavy. On the other hand, at $Q^2 \gg 10^2$ GeV$^2$, 
where the bottom contribution becomes visible,  
$s,c$ and even $b$ can be safely considered light quarks and
only $t$ is really heavy. How should one treat ``heavy'' flavours
in the large--$Q^2$ limit, where collinear singularities emerge ?
And how should one deal with the coexistence of ``light''
and ``heavy'' quarks in the same QCD processes, 
which is typical of charged--current DIS
and represents a further source of ambiguity ?

Various schemes for a NLO QCD analysis of heavy quark
contributions to electroweak structure functions are available
on the market,
but a 
detailed comparison of their predictions and 
their stability against changes of the factorization scale 
is still lacking. In this paper we want to bridge this gap, 
bringing to completion a program initiated in 
\cite{BGNPZ5,BDG}\footnote{A preprint has recently appeared \cite{Kramer}
 which partly
overlaps the study carried out in Ref.~\cite{BDG} and here. However  
a different perspective is adopted in \cite{Kramer}.}.
We shall offer a panoramic and detailed view of the heavy flavour
sectors of $F_2$ and $F_L$, with their dependence on $x$, $Q^2$, 
and in particular on the NLO scheme and the factorization scale. 
Results for $F_3$ have been presented elsewhere 
\cite{BDG}.

Let us start from the 
QCD factorization formula for the $q \bar q'$ contribution to the 
charged--current nucleon structure functions $F_2$ and $F_L$ ($q=c,t; 
\,\, \bar q'=\bar s, \bar b$), that can be formally written as
($\otimes$ means convolution)
\be
F^{qq'}_{2,L}(Q^2) = 
\sum_a f_a(\mu^2) \otimes \hat F^{a}_{2,L}(\mu^2, Q^2)\,.
\label{1}
\ee
Here the sum is made over all parton species, $\hat F^{a}_{2,L}$ are
the structure functions for $W$ scattering on parton $a$, and 
$\mu^2$ is the factorization scale. 

At order $\alpha_s^0$ only quarks and antiquarks contribute to
the sum in (\ref{1})
and $\hat F^{a}_{2,L}$
are proportional to the cross sections for the excitations
$W^{+}q' \rightarrow q, \,\,
W^{+}\bar q \rightarrow \bar q'$, which are essentially 
delta functions of the slow rescaling variables \cite{libri}. 
For $cs$ one has, 
{\it e.g.} 
\be
F^{cs}_2(x) = 2 \, \xi \, [ s(\xi) + \bar c (x) ]\,, 
\label{2}
\ee
where $\xi = (1 + m_c^2/Q^2)\, x$. The longitudinal structure 
function, although nonvanishing, 
is suppressed by a factor $m_c^2/Q^2$ ($m_c$ is the mass of the charmed
quark).
Notice that eq.~(\ref{2}) is a parton model formula: the only $Q^2$
dependence is of a kinematical nature. 

The slow rescaling (SR) procedure consists in using eq.~(\ref{2})
with a $Q^2$ dependence {\it \`a la} Altarelli--Parisi 
of the parton distributions. For $cs$
\be
{\rm SR:} \;\;\;\;\;
F^{cs}_2(x,Q^2) = 2 \, \xi \, [ s(\xi,Q^2) + \bar c (x,Q^2) ]\,. 
\label{3}
\ee
We recall that in some experimental papers \cite{CCFR1,CCFR2} 
a ``leading order analysis'' of data is performed, which 
is in fact the slow
rescaling analysis sketched above.

At order $\alpha_s$ the main contribution to the $qq'$ component
of the structure functions is 
given by the $W$--gluon fusion (GF) term,  
which
reads 
\be
{\rm GF:}\;\;\;\;\;
F_{2,L}^{qq'}(x,Q^2) = 
\frac{\alpha_s(\mu^2)}{2 \pi}  \, 
\int_{ax}^{1} 
\frac{{\rm d}y}{y} \, y \, g (y, \mu^2) \, 
\hat F^{g}_{2,L} \left(\frac{x}{y}, Q^2 \right)\,,
\label{a0}
\ee
where $a = 1 + (m^2 + m'^2)/Q^2$ and 
$g(y,\mu^2)$ is the gluon density at the factorization scale. 
The explicit expressions for $\hat F^{g}_{2}$ and $\hat F^{g}_{L}$
are, in the limit $m' \rightarrow 0$ \cite{WGR,GR}\footnote{For the 
sake of simplicity, we  
shall assume in our formulas that both the physical scale $Q^2$ and
the squared mass $m^2$ of the quark $q$ are much 
larger than the squared mass $m'^2$
of the quark $q'$; this is of course always true for the $cs$ sector, 
and is true also for the $bt$ sector at $Q^2 \gg 10^2$GeV$^2$. In the 
calculations the appropriate masses are used and no approximation is 
made.}
\bq
\hat F^{g}_2(z,Q^2) &=& 2z\, \left \{ 
  (1- \frac{m^2}{\hat s}   )
\, \left [   8z (1-z) - 1 - \frac{m^2}{Q^2}\, (1 -
 \frac{m^2}{Q^2} ) \, z(1-z) \right ] \right. \nonumber \\
&+& \left. ( 1 + \frac{m^2}{Q^2})\, {\cal P}_{qg}(\tilde z) \,  
 \left [ \log{ \frac{\hat s}{m^2}} + 
\log{ \frac{(\hat s - m^2)^2}{\hat s \, m'^2} } \right ]
+ 6z (1-2z) \, \frac{m^2}{Q^2} \, \log{ \frac{\hat s}{m^2}}
 \right \}\,\,, 
\label{4}
\eq  
\bq
\hat F^{g}_L(z,Q^2) &=& 2z\, \left \{ 
  (1- \frac{m^2}{\hat s}   )
\, \left [   4z (1-z)  - \frac{m^2}{Q^2}\, (1 -
 \frac{m^2}{Q^2} ) \, 2z(1-z) \right ] \right. \nonumber \\
&+& \left. \frac{m^2}{Q^2}\, {\cal P}_{qg}(\tilde z) \,  
 \left [ \log{ \frac{\hat s}{m^2}} + 
\log{ \frac{(\hat s - m^2)^2}{\hat s \, m'^2} } \right ]
+ 4z (1-2z) \, \frac{m^2}{Q^2} \, \log{ \frac{\hat s}{m^2}}
 \right \}\,\,, 
\label{5}
\eq  
In eqs.~(\ref{4},\ref{5}) $\hat s = Q^2 (1-z)/z$ and 
${\cal P}_{qg}$ is the usual $g \rightarrow q\bar q$
splitting function expressed in terms of the rescaled
variable $\tilde z = z (1+ m^2/Q^2)$.

When the quark $q'$ is a light flavour, which means either
$m'=0$ or $Q^2 \gg m'^2$, the logarithm $L(m',m) = 
\log{ [(\hat s - m^2)^2/\hat s \, m'^2] }$ develops a 
(collinear) singularity. Same considerations apply also to 
the logarithm $L(m,m') = 
\log { (\hat s/m^2)}$, when  $Q^2 \gg m^2$. There are 
various methods to subtract (or to deal with) 
such collinear divergences.

Let us focus for the moment on the charm--strange sector. 
We shall discuss three different prescriptions. 
Two of them
are variations of the so--called ``fixed flavour scheme'' 
(FFS) \cite{GR}. In this 
scheme the charmed quark is not considered to be a parton, 
{\it i.e.} an internal constituent of the nucleon, and its 
${\cal O}(\alpha_s^0)$ excitation diagram is omitted. 
We call FFS(a) and FFS(b) 
the two realizations of the fixed flavour scheme
that we shall consider in the following.
They differ for the treatment of the strange quark: in FFS(a) 
strangeness is
considered as a heavy flavour, with a mass $m_s \simeq 0.3-0.5$ GeV$^2$; 
in FFS(b) the strange quark is viewed as a massless parton.
The third prescription is the ``variable flavour scheme'' (VFS) 
\cite{Tung} 
in which  
charm is considered as a partonic constituent of the nucleon at 
large $Q^2$ and $\mu^2$ ($Q^2, \mu^2 \gg m_c^2$) and 
the ${\cal O}(\alpha_s^0)$ charm excitation diagram is taken into 
account, with a proper subtraction term. 

In the FFS(a) scheme one treats the strange quark (but not the charmed 
one) as a parton, setting $m'$ to zero and subtracting in 
eqs.~(\ref{4},\ref{5}) the terms proportional to $\log{Q^2/m'^2}$, which
are embodied in the Altarelli--Parisi evolution of the strange
distribution. In explicit form, the FFS(a) prescription reads
\be
{\rm FFS(a):}\;\;\;\;\;
F_{2,L}^{cs}(x,Q^2) = 2\, \xi \, s(\xi, Q^2) + 
\frac{\alpha_s(\mu^2)}{2 \pi}  \, 
\int_{ax}^{1} 
\frac{{\rm d}y}{y} \, y \, g (y, \mu^2) \, 
\tilde F^{g}_{2,L} \left(\frac{x}{y}, Q^2 \right)\,,
\label{6}
\ee
where $\tilde F^{g}_{2,L}$ means $\hat F^{g}_{2,L}$, 
eqs.~(\ref{4},\ref{5}), 
with 
the terms proportional to $\log{Q^2/m'^2}$ subtracted off. 

In the FFS(b) scheme eq.~(\ref{a0}) is the only contribution 
to $F_{2,L}^{cs}$. A small but finite mass for the strange quark
is retained and the full dependence on $m$ and
$m'$ in the gluonic structure functions (\ref{4}) and (\ref{5}). 
Of course no singularity appears. Explicitly one has
\be
{\rm FFS(b):}\;\;\;\;\;
F_{2,L}^{cs}(x,Q^2) = 
\frac{\alpha_s(\mu^2)}{2 \pi}  \, 
\int_{ax}^{1} 
\frac{{\rm d}y}{y} \, y \, g (y, \mu^2) \, 
\hat F^{g}_{2,L} \left(\frac{x}{y}, Q^2 \right)\,,
\label{7}
\ee
with $m \equiv m_c \neq 0$ and $m' \equiv m_s \neq 0$. 

In the fixed flavour schemes the number $n_f$ 
of active flavours is set to 3, irrespective of the scale of 
$\alpha_s$. 

In the VFS scheme the strange quark is a massless
parton and the charmed quark too is treated as a parton at large
$Q^2$. Both the strange and the charm ${\cal O}(\alpha_s^0)$ excitation 
diagrams are considered but the $Q^2$ dependence of the parton 
distributions is kept in the kernel of the ${\cal O}(\alpha_s^1)$
factorization integral. The collinear singularities are subtracted 
out by setting 
\be
L(m,m') \rightarrow \log{\frac{\hat s}{\mu^2}} \,,\;\;\;
L(m',m) \rightarrow \log { \frac{(\hat s - m^2)^2}{\hat s \, \mu^2} }\,,
\label{8}
\ee
where the subtraction scale 
is customarily taken to be equal to the factorization scale.
Thus the VFS prescription is 
\bq
{\rm VFS:}\;\;\;\;\;
F_{2,L}^{cs}(x,Q^2) &=& 2\, \xi \, [s(\xi, \mu^2) + 
c(x,\mu^2)] \nonumber \\
&+& 
\frac{\alpha_s(\mu^2)}{2 \pi}  \, 
\int_{ax}^{1} 
\frac{{\rm d}y}{y} \, y \, g (y, \mu^2) \, 
\bar F_{2,L}^{g} \left(\frac{x}{y}, Q^2 \right)\,,
\label{9}
\eq
where $\bar F_{2,L}^{g}$ means the 
gluonic structure functions,
eqs.~(\ref{4},\ref{5}), with the two subtractions (\ref{8}). Notice also
that in this scheme the number of active flavours is 4. 

The bottom--top sector is treated along similar lines, but, 
out of the three
approaches that we have just outlined, only  
FFS(b) and VFS survive.

What experimentalists do in their analyses is to use one of the
methods presented above (including the slow rescaling prescription), 
choose a factorization scale and extract the charm and strange 
(or bottom and top) densities
from the measured 
structure functions. It is clear that there at least
two sources of uncertainty in this procedure: one coming from the 
choice of the scheme, the other coming from the choice of $\mu^2$. 
The strategy adopted here reverses
the experimental procedure: we use a set of 
parton distributions provided by a global fit and study 
the dependence on $Q^2$, $\mu^2$ and on the scheme 
of the charged--current
structure functions in the heavy quark sector. 

In our calculations we use the MRS-A fit of parton densities 
\cite{MRS2}. For consistency we adopt the same set of quark masses
of Ref.~\cite{MRS2}, that is $m_c^2 = 2.7$ GeV$^2$, 
$m_b^2 = 30$ GeV$^2$. We use also $m_t= 180$ GeV and, for the 
FFS(b), $m_s = 0.3$ GeV. 
As for the
factorization scale, authors who adhere to different NLO schemes
give different suggestions: in the fixed flavour scheme of 
Ref.~\cite{GRS}, a value $\mu^2 = m_c^2$ 
is preferred, whereas 
in the variable flavour scheme of Ref.~\cite{Tung} a value 
which asymptotically becomes of order of $Q^2$ 
is suggested, and $\mu^2= Q^2/2$ is practically used.  
For homogeneity we decided to choose the same value for 
all schemes, thus we take $\mu^2= Q^2/2$
for the calculations presented in Figs.~1-4 
and postpone the illustration of the $\mu^2$ dependence to Fig.~5.

In Fig.~1 we present $F_2^{cs}$ 
at two different values ($Q^2 = 25, 100$ GeV$^2$). The results
of all prescriptions [FFS(a), FFS(b), VFS, SR] are shown. 
Consider, for instance, $F_2^{cs}$ at $Q^2 = 25$ GeV$^2$. 
If we take the 
difference between the results of the various prescriptions as 
a measure of the theoretical uncertainty arising from the choice 
of the scheme, this uncertainty amounts to $ \pm 3 \%$ at $x = 
10^{-3}$, and to $\pm  20 \%$ at $x = 10^{-1}$. 
We have checked that the 
three NLO schemes give largely different predictions
at
lower $x$ ($x < 10^{-4}$); however a correct analysis of this
region requires the use of the $k_t$ factorization method, 
which is beyond
the purposes of this work. A forthcoming paper is specifically 
devoted to this
issue \cite{BDG2}.
Note that 
the slow rescaling curve almost coincides with the VFS result. On the
other hand, the largest discrepancy occurs between the FFS(a) and 
the FFS(b) predictions, due to the ${\cal O}(\alpha_s^0)$ term
which is absent in FFS(b).
We mention that the FFS(b) result
for $cs$ is affected by an additional (and large) 
uncertainty coming from 
the choice of $m_s$. This makes FFS(b) intrinsically unreliable. 

If we confine ourselves to VFS and FFS(a), the theoretical 
uncertainty is $\pm (3-7) \%$ over the whole $x$ range at 
$Q^2 = 25$ GeV$^2$ and gets
smaller with increasing $Q^2$, being $\pm (2-4) \%$ at $Q^2 =
100$ GeV$^2$. Thus it will be important to have data 
on $F_2^{cs}$ at larger $Q^2$. 
 
In Fig.~2 the curves for 
the longitudinal charm--strange structure function $F_L^{cs}$ 
are displayed. The FFS(a) and FFS(b) results are much closer 
to each other in this case than in the $F_2^{cs}$ case. The theoretical
uncertainty amounts now to $ \pm 10 \%$
at $Q^2 = 25$ GeV$^2$ over the whole $x$ range. 
Obviously the SR curve for $F_L^{cs}$ is much lower, being 
suppressed by a factor $m_c^2/Q^2$. The slow rescaling approach
is expected to be unreliable when the longitudinal structure function 
dominates. The relevance of the longitudinal 
contribution is quantified by the ratio $\rho \equiv F_L/F_2$, which
is a measure of the violation of the Callan--Gross relation. 
In the $cs$ sector we find, in the FFS(a) scheme, $\rho^{cs}
\simeq 0.30$ at $x = 10^{-3} \div 10^{-2}$. 

The breaking of the Callan--Gross relation is stronger
in the $bt$ sector.
In Fig.~3 we show the SR, FFS(b) and VFS predictions 
for $F_2^{bt}$ at two $Q^2$ values ($Q^2 = 10^3, 10^4$ GeV$^2$). 
The difference between FFS and VFS amounts to $ \sim 10 \%$, whereas
the discrepancy between slow rescaling and NLO schemes is more
dramatic (up to $ \sim 70 \%$). We do not plot $F_L^{bt}$, limiting
ourselves to mention that it dominates $F_2^{bt}$. We find in fact
$\rho^{bt} (x=10^{-3}) = 0.97$ ({\it i.e.} $F_L^{bt} \simeq F_2^{bt}$)
at $Q^2 = 10^3$ GeV$^2$, and $\rho^{bt} (x=10^{-3}) = 0.79$ 
at $Q^2 = 10^4$ GeV$^2$ (these values are almost scheme independent). 

For completeness we show in Fig.~4 the $Q^2$ dependence of the 
structure functions $F_2^{cs}$ and $F_L^{cs}$ in two schemes, 
VFS and FFS(a). The trend is similar in the two cases.

Now, is there a way to prefer a scheme to another ? A practical 
criterion is to look at the dependence of the results on the 
factorization scale $\mu^2$. In Fig.~5 we present the $\mu^2$
dependence of $F_2^{cs}$ at $Q^2 = 25$ GeV$^2$ in each scheme.
It can be easily seen  that VFS and FFS(a) are more stable 
against changes in $\mu^2$. A slight preference should be 
attributed to the latter scheme, which gives a practically 
constant prediction in the range $\mu^2 = (5 - 100)$ GeV$^2$, whereas
in the same range the VFS result changes by $ \sim 25 \%$ at
$x = 10^{-3}$, and by $ \sim 15 \%$ at $x = 0.1$. 
We have checked that also at $Q^2 = 100$ GeV$^2$ 
FFS(a) is more stable than VFS in the range $\mu^2 = (5 - 1000)$ 
GeV$^2$. The main variation of VFS occurs for small $\mu^2$ values, 
so that if VFS is used a $\mu^2 \gsim Q^2/2$ is preferable.

We remark here that in the next--to--leading 
CCFR analysis of dimuon data \cite{CCFR2}, which resorts to the 
VFS scheme, the factorization scale is chosen to be equal 
to twice
the maximum available transverse momentum of the final charm quark, 
$\mu = 2 p_{\perp}^{max}$.
At $x= 10^{-2}$ this scale is about two orders of magnitude
larger than the CCFR average
$Q^2$, which is 22 GeV$^2$. The quoted 
uncertainty on the strange distribution
due to the arbitrariness of the choice of $\mu^2$ is about
$5 \%$ and is obtained by varying $\mu$ between $p_{\perp}^{max}$
and $3 p_{\perp}^{max}$. Our results show that the CCFR Collaboration
underestimates the 
theoretical uncertainties of its determination. 

Let us summarize our work. In this paper we provided an $(x,Q^2,\mu^2)$ 
map of the heavy quark contributions to charged current structure 
functions in some customary QCD schemes. The best way to choose
the optimal scheme and factorization scale is to study the perturbative
stability, which is possible only by pushing at least 
to order $\alpha_s^2$
the knowledge of the 
factorization kernels. Some partial results
heve been obtained in the electromagnetic case \cite{GRS,Olness}, 
but a similar treatment of the weak DIS is still lacking and 
will be object of a future investigation. However some conclusions 
can be already drawn. While the slow rescaling mechanism should
be definitely abandoned -- at least when longitudinal contributions 
are present (it may nevertheless be useful 
for the analysis of purely transverse
structure functions such as $xF_3$ \cite{BDG}) --,
it seems that, in the charm-strange
sector, 
the fixed flavour
scheme with subtraction for the strange quark -- 
the scheme we have called FFS(a) -- is preferable due to its greater
stability. In any case, the uncertainties coming from the choice
of the scheme and of the factorization scale add up to some tens 
per cent, and 
require a better consideration both on the experimental 
and on the phenomenological side.  
More data, especially at larger $Q^2$, are needed in order to achieve 
a well--established knowledge of the strange and charm distributions.

\pagebreak

\pagebreak

\begin{center}
{\Large \bf Figure Captions}
\end{center}

\vspace{1cm}

\begin{itemize}

\item[Fig.~1]
$F_2^{cs}$ at $Q^2 = 25, ~100$ GeV$^2$. The solid line corresponds to the
VFS result [see the text], the dot-dashed one to FFS(a), 
the dashed one to FFS(b), the dotted one to SR. 

\item[Fig.2]
$F_L^{cs}$ at $Q^2 = 25, ~100$ GeV$^2$. 
The curves are labelled as 
in Fig.~1.

\item[Fig.3]
$F_2^{bt}$ at $Q^2 = 10^3, ~10^4$ GeV$^2$. 
The curves are labelled as 
in Fig.~1.

\item[Fig.4]
The $Q^2$ dependence of $F_2^{cs}$ and $F_L^{cs}$. 
Only the VFS prediction (solid curve) and the FFS(a) 
(dot-dashed) prediction are shown.

\item[Fig.5]
The dependence of $F_2^{cs}$ on the factorization 
scale $\mu^2$ at $x= 0.001, ~0.01 $ and $0.1$
and $Q^2 = 25$ GeV$^2$.
The curves are labelled as in Fig.~1.

\end{itemize}

\end{document}